# Photometric Observations and Period Analysis of an SU UMa-type Dwarf Nova, MASTER OT J004527.52+503213.8

**Sena A. Matsui** (ORCID 0000-0003-4249-0792)
*Division of Particle and Astrophysical Science, Nagoya University, Furo-cho, Chikusa-ku, Nagoya 464–8602, Japan*
*Department of Biosphere-Geosphere Science, Okayama University of Science, 1-1 Ridaicho, Kita-ku, Okayama 700–0005, Japan; sena.ma@nagoya-u.jp*

**Tsutomu T. Takeuchi** (ORCID 0000-0001-8416-7673)
*Division of Particle and Astrophysical Science, Nagoya University, Furo-cho, Chikusa-ku, Nagoya 464–8602, Japan*
*The Research Center for Statistical Machine Learning, the Institute of Statistical Mathematics, 10-3 Midori-cho, Tachikawa, Tokyo 190—8562, Japan; takeuchi.tsutomu.g8@f.mail.nagoya-u.ac.jp*

**Kai T. Kono**
*Division of Particle and Astrophysical Science, Nagoya University, Furo-cho, Chikusa-ku, Nagoya 464–8602, Japan; kono.kai@c.mbox.nagoya-u.ac.jp*

**Suchetha Cooray** (ORCID 0000-0002-9217-1696)
*Division of Particle and Astrophysical Science, Nagoya University, Furo-cho, Chikusa-ku, Nagoya 464–8602, Japan*
*Research Fellow of the Japan Society for the Promotion of Science (DC1); cooray@nagoya-u.jp*

**Abstract** MASTER OT J004527.52+503213.8 (hereafter MASTER J004527) is a dwarf nova discovered by the MASTER project in 2013. At 18:20 UTC on 24 October 2020, brightening of this object was reported to vsnet-alert (24843 by Denisenko). This was the second report of a superoutburst after its discovery. Photometric observations were made using the 23.5-cm Schmidt-Cassegrain telescope at Okayama University of Science observatory soon after the alert through 4 November 2020. In this work, we present the photometric data from our observation, and the analysis of the light curves of MASTER J004527 during the 2020 outburst. We propose a method to determine the period of superhumps by polynomial fitting, which can be applied to a light curve with many missing data. In addition to our own data, we incorporate other all sky survey data of the outburst to better understand the properties of the superhumps. Based on our observations, we conclude that MASTER J004527 is an SU UMa-type dwarf nova, since no early superhumps occurred.

### 1. Introduction

Cataclysmic variables (CVs) are close binary systems, consisting of a white dwarf primary star and a late-type secondary star. The characteristic property of CVs is their rapid increase of luminosity. Dwarf novae (DNe) are one of the subclasses of CVs. There are three types of CVs: U Gem, Z Cam, and SU UMa. SU UMa-type DNe are further classified into three subtypes: SU UMa, WZ Sge, and ER UMa. Detailed information on the DN classification can be found in, e.g., La Dous (1994) and Osaki (1996).

In the subdivision of SU UMa-type dwarf novae, the definition of the WZ Sge type is the observation of early superhumps. The early superhumps are small amplitude fluctuations of 0.1 to 0.5 magnitude that appear for about a week after the maximum magnitude. They are thought to be the result of tidal instability caused when the outer disk reaches a 3:1 resonance radius during an outburst (e.g., Osaki 1989; Hirose and Osaki 1990).

On 25 October 2020 (JST), an outburst of a DN, MASTER OT J004527.52+503213.8 (hereafter MASTER J004527), was reported on VSNET (Denisenko 2020). Denisenko *et al.* (2013) mentioned that, based on the blue color and outburst amplitude, MASTER J004527 was most likely a WZ Sge in superoutburst. Kato (2015) and AAVSO VSX labeled it as SU UMa-type.

During this outburst, reported on vsnet as a second confirmed superoutburst, MASTER J004527 increased its brightness by up to $\simeq 13$ mag. The increased brightness was sufficient to be observed by the 23.5-cm Schmidt-Cassegrain telescope at the observatory of Okayama University of Science in Japan. We conducted a photometric observation of MASTER J004527 with this telescope and obtained the light curve from 25 October through 3 November 2020 (JST).

In this study, we present the estimated parameters from the analysis of the light curve. Further, we compared our own observation data with other survey data from public databases. The data from these surveys provide information on the global characteristics of the light curve. We referred to the data obtained by the All-Sky Automated Survey for Supernovae (ASAS-SN) and the Zwicky Transient Facility Survey (ZTF). The All-Sky Automated Survey for Supernovae (here after "ASAS-SN") project surveys automatically the sky almost every night with 24 telescopes located all over the world. The Zwicky Transient Facility (here after "ZTF") is a survey of the wide field astronomy with the Samuel Oschin Telescope at Palomar Observatory in California, United States.

We present an analysis to support the classification of MASTER J004527.

This paper is organized as follows. In section 2 we introduce all the datasets we used for this study. We explained the analysis methods in section 3. Section 4 presents the observed results. We discuss some physical interpretations of the results in section 5. Section 6 is devoted to our conclusion. We explained some detailed information on the observation and data analysis in Appendix A.

## 2. Data

### 2.1. Target object

MASTER J004527 was discovered by the MASTER project in 2013 during its outburst. It had become 12.53 mag in Clear filter at the time of discovery on 17.668 September 2020 UTC, which was reported by Denisenko (2013). MASTER J004527 is located at R.A. $00^h \, 5^m \, 27.54 \pm 0.18^s$, Dec. $+50° \, 32' \, 15.18 \pm 0.17"$ (J2000). The magnitudes of MASTER J004527 in the quiescent period are presented in Table 1.

### 2.2. Observation and data reduction

We performed photometry of the target object MASTER J004527 during the period from 25 Oct. 2020 to 4 Nov. 2020 (JST), with one of the facilities at the Observatory of Okayama University of Science, Japan. The telescope was a Schmidt-Cassegrain, with an aperture of 235 mm and a focal length of 1480 mm. We used a cooled CCD camera, SBIG ST-9XE with 512 × 512 pixels (pixel size 20 × 20 μm). In addition, we used a Clear filter, and the exposure time was 60 s throughout this observation. The data were reduced with a standard procedure by using AstroImageJ (ver. 3.2.0) developed by Collins *et al.* (2017). Our observation log is given in Table A1, shown in Part A1 of Appendix A.

### 2.3. Light curves

We present the light curve of MASTER J004527 in the whole observation period in Figure 1. Abscissa is the Julian day, subtracted with a constant so that the light curve starts from zero. Ordinate represents Δmag of MASTER J004527 compared to the standard star in the same field of view. The Δmag is defined as a relative magnitude between the comparison star TYC 3257-553-1 (denoted by C3) and a target star. Since we used a small-aperture telescope with a Clear filter, we used Δmag for the discussion on the photometry. We chose comparison stars in the neighborhood of the target MASTER J004527. We denote the magnitude of MASTER J004527 as T1. The comparison stars are TYC 3270-1038-1 located at R.A. $00^h 45^m 13.6689 \pm 0.0197^s$, Dec. $+50° 30' 40.2192 \pm 0.0195''$ (J2000) and TYC 3257-553-1 located at R.A. $00^h 44^m 52.0981 \pm 0.0135^s$, Dec. $+50° 28' 11.6309 \pm 0.0134''$ (J2000). Both stars have been confirmed that they are not variable stars. The magnitude of TYC 3270-1038-1 is $11.545754 \pm 0.002761$ in the Gaia G-band (Gaia Collaboration 2020) and $11.51 \pm 0.09$ in the V-band (Høg *et al.* 2000). The magnitude of TYC 3257-553-1 is $11.947890 \pm 0.002763$ in the Gaia G-band (Gaia Collaboration 2020) and $12.10 \pm 0.17$ in the V-band (Høg *et al* 2000). We denote the magnitudes of TYC 3270-1038-1 and TYC 3257-553-1 as C2 and C3, respectively. The relative magnitude Δmag of the target star is T1 – C3. We also estimated Δ mag of the comparison star TYC 3270-1038-1, C2 – C3, to examine the stability of the photometry. We present T1 – C3 and C2 – C3 in Figure 1.

Table 1. Magnitude in the quiescent period.

| Band | Magnitude (mag) | Reference |
|---|---|---|
| R | 19.9 | Monet *et al.* (2003) |
| B | 19.7 | Monet *et al.* (2003) |
| Gaia G[a] | 18.940004 ± 0.006565 | Gaia Collab. (2020) |

[a] For the definition of Gaia G-band, see
 https://www.cosmos.esa.int/web/gaia/edr3-passbands.

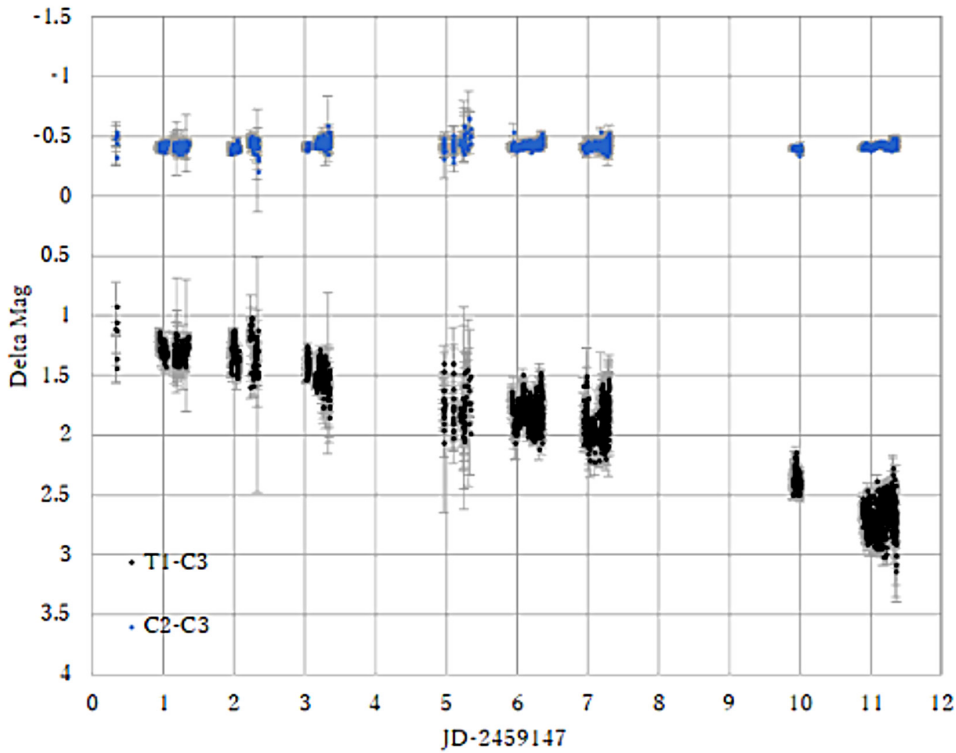

Figure 1. A light curve of MASTER OT J004527.52+503213.8 (MASTER J004527) during the whole observation period. Relative magnitude obtained by the standard star photometry is shown. The abscissa is the Julian date subtracted with a constant so that the light curve starts from zero. The ordinate represents the Δ mag, i.e., the magnitude obtained by subtracting a magnitude of the comparison star TYC 3257-553-1 (denoted by C3) from that of a target star. The black symbols are the magnitude of the target star (T1) minus the magnitude of the comparison star C3, while the blue symbols are the magnitude of comparison star TYC 3270-1038-1 (denoted by C2) minus C3. See main text for the details.

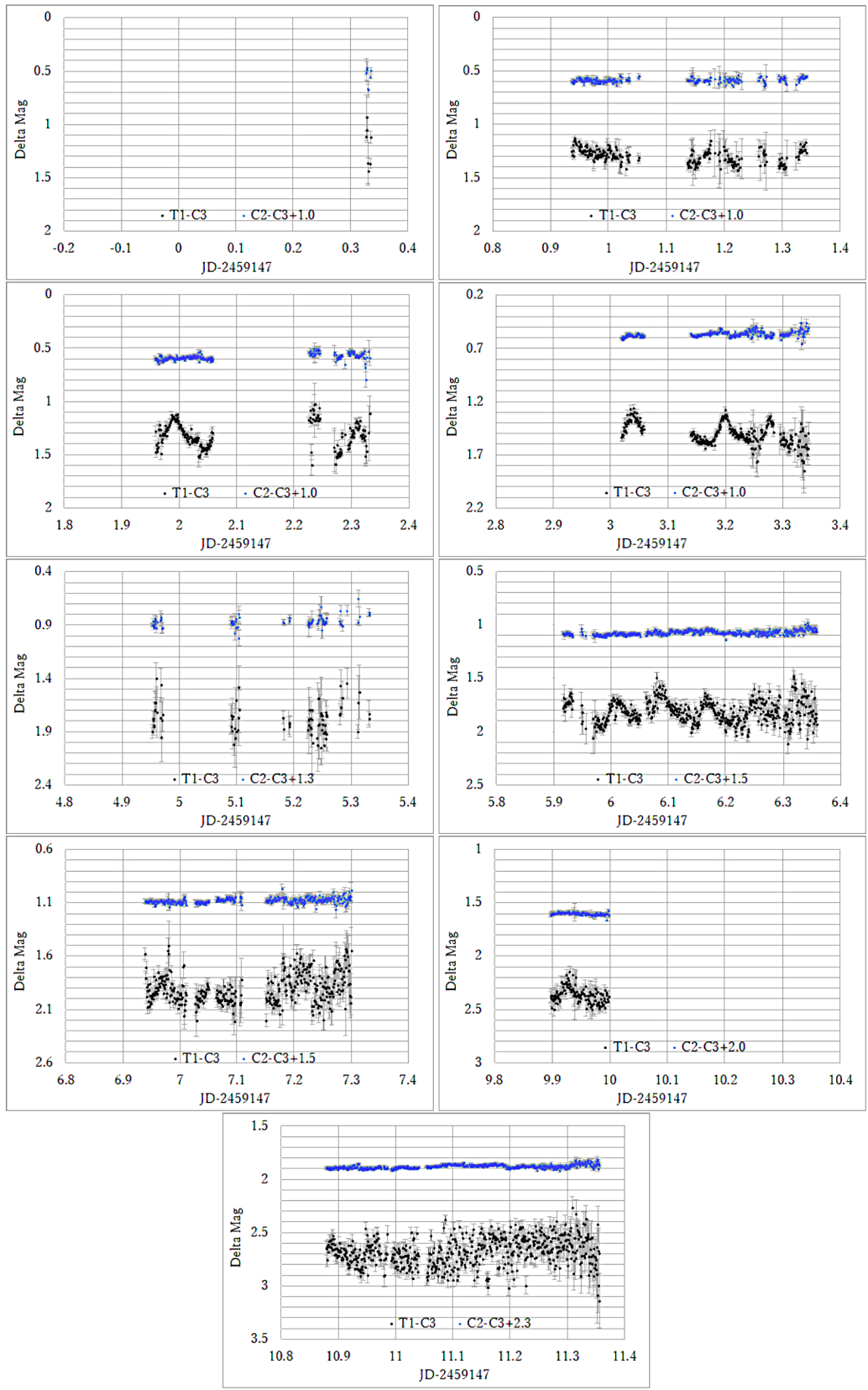

Figure 2. Detailed light curves of MASTER J004527. Symbols and format are the same as Figure 1.

In Figure 1, we observe a dimming of $\simeq 1.5$ mag in ten days. A light curve for each observation day is shown in Figure 2[1]. The target star is diminishing after the outburst, while the relative magnitude of the comparison star C2 – C3 stays constant. Therefore, in Figue 2, a constant is added to Δmag in each panel, to avoid C2 — C3 values to be too far from T1 — C3, to make the comparison easily. The specific values of the constant are specified in each panel. For example, the top two panels show C2 – C3 + 1.0.

## 3. Method

In this work, we estimated the period of superhumps by fitting polynomials to the light curve of each hump and estimating the time of extrema (peaks). We adopted this method mainly because the error is significantly large for a part of the data, and its analysis is straightforward. Fitting Errors were calculated by the standard Jackknife resampling method (Efron 1982).

We should note that the phase dispersion minimization method (PDM; Stellingwerf 1978) has been used as a standard procedure for the period analysis of CVs (e.g., Kennedy *et al.* 2016; Tanabe *et al.* 2018). However, its performance is guaranteed only for continuous data. The observed data in this study were not globally contiguous in time, and the PDM method was not suitable for this analysis. This is the reason why we adopted the polynomial fitting method for the period analysis, instead of the PDM. For comparison, we performed a PDM analysis for each continuous portion of the light curve. The results are shown in Part A3 of Appendix A.

### 3.1. Estimation of the time of hump maxima
#### 3.1.1. Time of hump maxima

To estimate the timing of peaks of the superhumps, we fitted second- and third-order polynomials to each hump in the light curves. Since what we should find is only the timing of a peak, we do not have to consider higher-order polynomials. We determined which order is more appropriate to describe the hump, we evaluated the Akaike information criterion (AIC; Akaike 1974) and Bayesian information criterion (BIC; Schwarz 1978). Formulation of AIC and BIC is provided in Part A2 of Appendix A.

The fitting results are presented in Figures 3 through 5, and the obtained peak times are tabulated in Table 2. In Table 2, we summarize the information on the peaks of humps estimated by the polynomial fitting. For some humps, the AIC and BIC suggest different conclusions and we cannot determine which order is better to fit. We adopted the second-order peak in such a case, since the error of the parameter estimation is smaller. The order of the selected fitting polynomial model k is also tabulated in Table 2.

As a first step, we start from a first-order approximation that the period is constant. We estimated the period of humps from the difference between the two detected peaks. For this, we assumed the following relation

$$O = T_0 + EP, \quad (1)$$

where O is an estimated time of a peak, $T_0$ is the time of the first peak that occurred on 25 Oct. 2020 (JST), and P is a period temporarily determined from the average for sequential peaks observed on 27 and 30 Oct. 2020 (JST).

By rearranging Equation 1, we have

$$E = \frac{O - T_0}{P}, \quad (2)$$

where $T_0$ is the time of the first peak obtained by observation, and P is a tentative period averaged over the difference between two successive peaks that could be observed. In this work, we adopted $T_0 = 1.9912$ and $P = 0.08058$ to estimate the epoch E. Since E should be an integer in principle, we round off E from Equation 2. We denote the rounded E as [E]. The precise period is then estimated from a scatter plot between [E] and O. The slope of the linear fit to the [E] – O relation yields the proper estimation of the period between the humps.

Table 2. Time of the detected maxima.

| Date (JST) | Time of maxima (JD–2459147) |
|---|---|
| 26 Oct 2020 | $1.9912 \pm 0.0004$ |
| 27 Oct 2020 | $3.0393 \pm 0.0006$ |
|  | $3.1996 \pm 0.0008$ |
|  | $3.2795 \pm 0.0007$ |
| 30 Oct 2020 | $6.0110 \pm 0.0005$ |
|  | $6.0870 \pm 0.0131$ |
|  | $6.1679 \pm 0.0005$ |
|  | $6.2534 \pm 0.0101$ |

---

[1] The machine-readable data are available from the following URL: https://drive.google.com/file/d/1Zwqui6r36J4RQmYP2dPqDIsWEewd1wsb/view?usp=sharing.

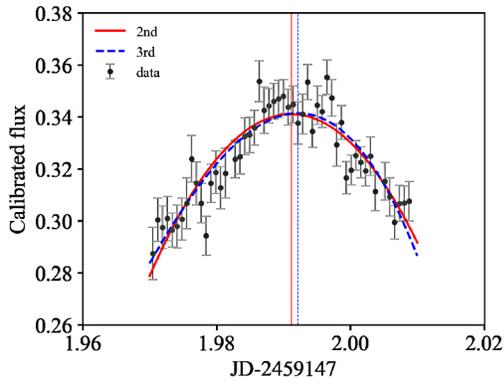

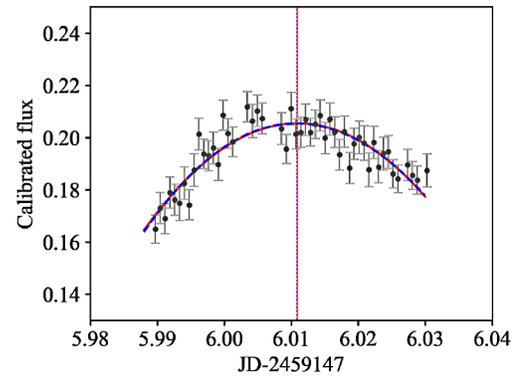

Figure 3. Parameter estimation of the time of hump maxima of MASTER J004527 on 26 Oct. 2020. The abscissa is the Julian date, and the ordinate is the flux of MASTER J004527 calibrated by the standard star in the FoV. The solid curve represents the second-order polynomial fit, and the dashed curve is the third-order fit. Vertical lines represent the estimated timing of the peak of the superhump. Detailed values related to the fit are tabulated in Table 4.

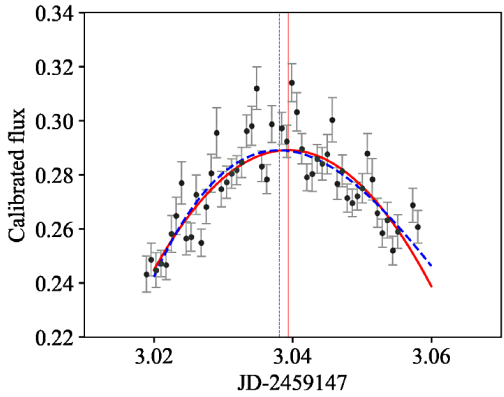

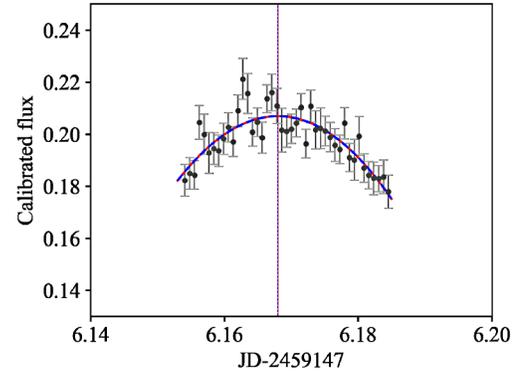

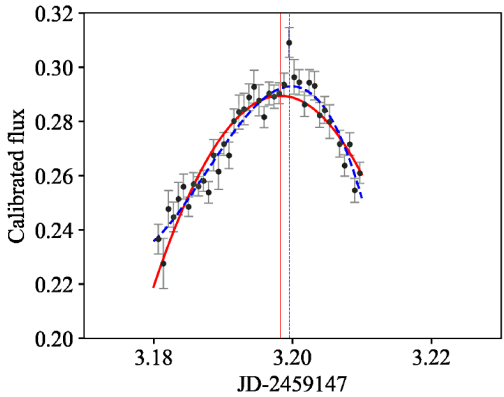

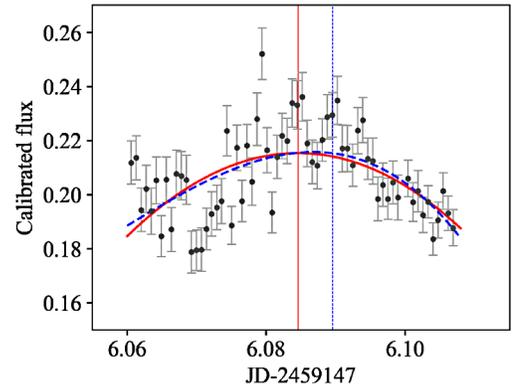

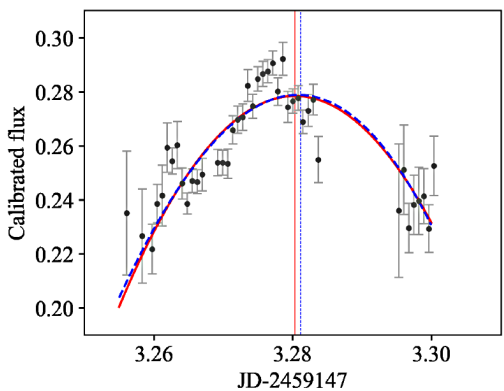

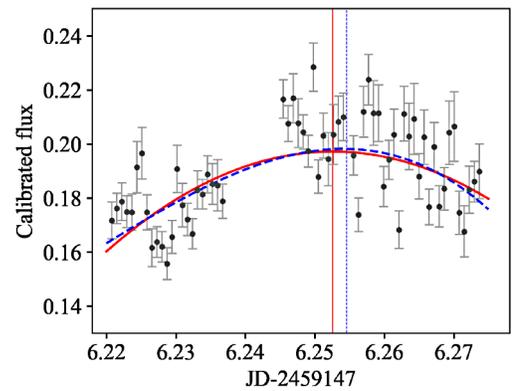

Figure 5. Same as Figure 3 but for the peaks on 30 Oct. 2020.

Figure 4. Same as Figure 3 but for the peaks on 27 Oct. 2020.

## 4. Results

4.1. Global behavior of the light curve

Figure 6 shows the light curves obtained from our observation and from all-sky surveys. We refer to the data obtained by the All-Sky Automated Survey for Supernovae (ASAS-SN) and the Zwicky Transient Facility Survey (ZTF). According to these observations, the plateau phase lasted about 12 days, during which MASTER J004527 dimmed by about 2 mag. We find no re-brightening after the superoutburst, which is typically observed in WZ Sge-type DNe. Furthermore, several small outbursts occurred after the superoutburst, according to the ASAS-SN data (Figure 7). We discuss this in more detail in section 5.2.

4.2. Peaks and periodic analysis

We then analyze the relation between E and O by a linear fitting. The period of the humps is estimated to be 0.08034 ± 0.00003 day, corresponding to 115.69 ± 0.05 min.

To examine the variation of the period of humps, we performed the so-called O–C diagram analysis. The name O–C stands for "Observed minus Calculated." It is expressed as deviations of phase in the cycle of variability. We followed the standard procedure for the analysis see, e.g., Sterken (2005). The period obtained from the linear fit is adopted as the value of the period P', and we calculated C as

$$C = T_0 + [E]P'. \quad (3)$$

We calculated O–C using the O obtained by observation and C obtained by calculation, and we made the O–C diagram (Figure 8).

The obtained quantities for the O–C analysis are listed in Table 3. Figure 8 is the O–C diagram of MASTER J004527. As described in section 3.2, the abscissa represents the rounded value of the Epoch, [E], and the ordinate is the Observation data minus the Calculation data. However, since the data points are too few in Figure 8, it is difficult to discuss variation of the period only with the current data.

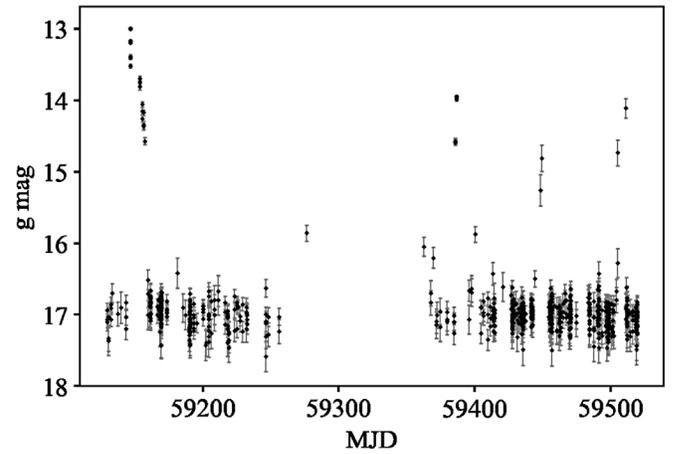

Figure 7. The light curve of MASTER J004527 for 391 days from MJD 59129 to MJD 59520 obtained by ASAS-SN. There are several outbursts without a hump after the superoutburst. The horizontal and vertical axes are the same as the upper panel in Figure 6.

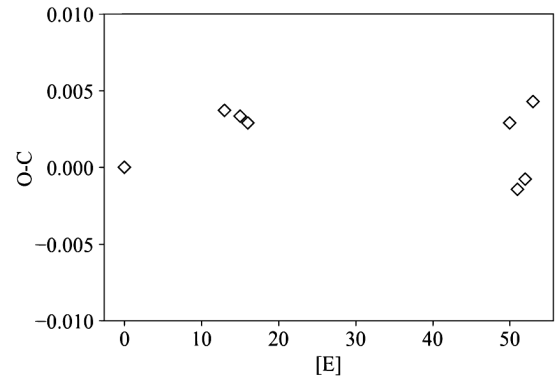

Figure 8. O–C diagram of MASTER J004527.

Table 3. Observed quantities for the O–C analysis.

| O | E | [E] | C | O–C |
|---|---|---|---|---|
| 1.9912 ± 0.0004 | 0 | 0 | 1.9912 | 0.0000 |
| 3.0393 ± 0.0006 | 13.01 | 13 | 3.0356 | 0.0037 |
| 3.1996 ± 0.0008 | 15.00 | 15 | 3.1963 | 0.0033 |
| 3.2795 ± 0.0007 | 15.99 | 16 | 3.2766 | 0.0029 |
| 6.0110 ± 0.0005 | 49.89 | 50 | 6.0081 | 0.0029 |
| 6.0870 ± 0.0131 | 50.83 | 51 | 6.0884 | –0.0014 |
| 6.1679 ± 0.0005 | 51.83 | 52 | 6.1688 | –0.0008 |
| 6.2534 ± 0.0101 | 52.89 | 53 | 6.2491 | 0.0043 |

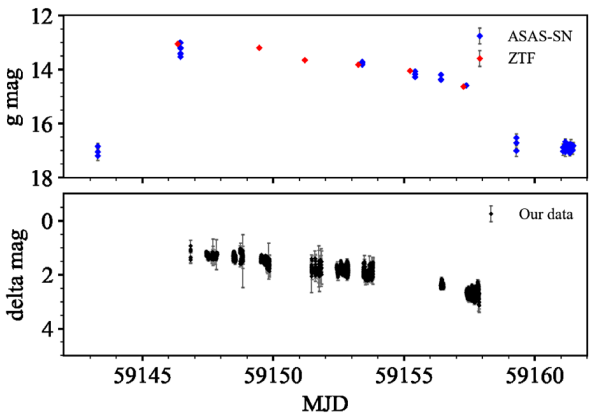

Figure 6. Light curves of the two all-sky surveys compared with our observed data. Upper panel: light curves from the ASAS-SN and the ZTF. Blue symbols represent the ASAS-SN data, while red ones are the ZTF data. The ordinate is in g magnitude. Lower panel: our observed light curve. The ordinate is expressed in Δmag, the difference between the measured brightness of MASTER J004527 and the standard star in the FoV. The abscissa is in MJD, an index of JD minus 2400000.5 days.

## 5. Discussion

### 5.1. Examination of the light curve of the superoutburst

In the subclassification of SU UMa-type dwarf novae, one of the definitions of the WZ Sge type is the so-called "early superhumps" feature. The early superhumps are small-amplitude fluctuations of 0.1 to 0.5 magnitude that appear for about a week after the maximum magnitude. They are thought to be the result of tidal instability caused when the outer disk reaches a 3:1 resonance radius during an outburst (e.g., Osaki 1989; Hirose and Osaki 1990). In our observation, there was no variation that could be considered early superhumps, though the data points are not enough to give a definitive conclusion.

Comparison of our data with the mainstream sky survey data from ASAS-SN and ZTF from Figure 7 clearly shows that the light curves are consistent with each other. This confirms the reliability of the data in this work. Although the precise date and time of the outburst cannot be determined, at least the outburst occurred at some moment during 3.15477 days between 59143.2778 MJD (the last observation before the outburst) and 59146.43257 MJD (when the outburst was detected). Therefore, our observations should have started within 3.54926 days after the outburst at most. The first superhump-like feature was detected at 59147.69086 MJD, which means that the superhumps were detected between 1.25829 and 4.41296 days after the outburst. This means that, since usually the early superhumps appear approximately one week after the maximum, it is not very plausible that we have missed the early superhumps associated with this superoutburst. Then, we conclude that the early superhumps may not have occurred. Early superhumps were also not reported in the 2013 superoutburst (Kato 2015).

### 5.2. Normal outbursts after the surperoutburst

The ASAS-SN data in Figure 7 show that since the superoutburst in 2020, a number of humpless outbursts with a smaller amplitude have been detected. It is highly probable that they are normal outbursts. Generally, WZ Sge-type dwarf novae do not have normal outbursts (see Patterson *et al.* 1981). This is consistent with the classification of MASTER J004527 as a SU UMa-type DN.

### 5.3. Classification of MASTER J004527

Now we consider the classification of MASTER J004527. As discussed above, the estimated period of superhumps of MASTER J004527 is about 116 min, strongly supporting that it should be classified as a typical SU UMa-type DN (period ~ 90–120 min). In comparison, the period of WZ Sge type objects is about 80 min (e.g., Tanabe *et al.* 2018). The measured superhump period is too long for MASTER J004527 to be classified as a WZ Sge-type object (e.g., Vogt 1980).

In contrast, Kato (2015) proposed a measure of how many magnitudes brighter the superhumps are when they appear compared to the quiescent magnitude, and many WZ Sge-type objects have amplitudes of 7 magnitudes or brighter. The exact magnitude is not known from this observation, but considering the all-sky surveys data, the amplitude is considered to be about $\simeq 4$ mag. Again, this suggests that it has the characteristics of a SU UMa-type DNe. Putting all discussions together, we conclude that MASTER J004527 is an SU UMa-type DN.

## 6. Summary

MASTER OT J004527.52+503213.8 (MASTER J004527) is a dwarf nova (DN) discovered by the MASTER project in 2013. This DN is considered to be an SU UMa-type. In this study, we present an analysis to support the classification of MASTER J004527. At 18:20 UTC on 24 Oct. 2020, brightening of this object was reported to vsnet-alert by Denisenko (2020). MASTER J004527 had brightened to ~13 mag during the superoutburst, enough to be detected by the 23.5-cm Schmidt-Cassegrain telescope at Okayama University of Science Observatory. We conducted a photometric observation of MASTER J004527 soon after the alert through 4 Nov. 2020 and obtained the light curve. We provide our own photometric data publicly online through the URL mentioned in section 2.

A comparison of this observation and other all-sky surveys has shown that early superhumps may not have occurred. This indicates that MASTER J004527 is an SU UMa-type DN. Although we could not prove the change in period with the current data, we can also consider that MASTER J004527 belongs to the SU UMa type based on the change in period observed during the 2013 outburst (Kato 2015). In addition, the observation of multiple normal outbursts further supports that MASTER is an SU UMa-type DN.

In this study, we applied a method to calculate the peak of superhumps from polynomial fitting, instead of the standard PDM. Since PDM can only handle continuous data, our method is more suitable to analyze data with many missing data, as in this study. Our method has another advantage in that is more intuitive and easier to understand. Further, more data-scientific approach can be done. We mention some possible methods in Part A3 of Appendix A, but we leave it as our future work.

## 7. Acknowledgements

This work has been supported by the Japan Society for the Promotion of Science (JSPS) Grants-in-Aid for Scientific Research (19H05076 and 21H01128). This work has also been supported in part by the Collaboration Funding of the Institute of Statistical Mathematics "New Development of the Studies on Galaxy Evolution with a Method of Data Science." SC is supported by the Japan Society for the Promotion of Science (JSPS) under Grant No. 21J23611. We thank Shiro Ikeda for discussing detailed data analysis methods, and Makoto Uemura, Daisaku Nogami, and Agnieszka Pollo for important suggestions in the discussion. We are sincerely grateful to the All-Sky Automated Survey for Supernovae (ASAS-SN) and the Zwicky Transient Facility Survey (ZTF) for providing the data of MASTER J004527.

# References

Akaike, H. 1974, *IEEE Trans. Automatic Control*, **19**, 716.
Banks, H., and Joyner, M. L. 2017, *Appl. Math. Lett.*, **74**, 33.
Collins, K. A., Kielkopf, J. F., Stassun, K. G., and Hessman, F. V. 2017, *Astron. J.*, **153**, 77.
Cooray, S., Takeuchi, T. T., Akahori, T., Miyashita, Y., Ideguchi, S., Takahashi, K., and Ichiki, K. 2021a, *Mon. Not. Roy. Astron. Soc.*, **500**, 5129.
Cooray, S., Takeuchi, T. T., Ideguchi, S., Akahori, T., Miyashita, Y., and Takahashi, K. 2021b, arXiv:2112.01444.
Cooray, S., Takeuchi, T. T., Yoda, M., and Sorai, K. 2020, *Publ. Astron. Soc. Japan*, **72**, 61.
Denisenko, D. 2013 vsnet-alert 16417 (September 19; http://ooruri.kusastro.kyoto-u.ac.jp/mailarchive/vsnet-alert/16417).
Denisenko, D. 2020, vsnet-alert 24843 (October 25; http://ooruri.kusastro.kyoto-u.ac.jp/mailarchive/vsnet-alert/24843).
Denisenko, D., *et al.* 2013, *Astron. Telegram*, No. 5399, 1.
Efron, B. 1982, *The Jackknife, the Bootstrap and Other Resampling Plans*, Society for Industrial and Applied Mathematics (SIAM), Philadelphia.
Gaia Collaboration. 2020, VizieR On-line Data Catalog: I/350, originally published in 2020A&A...649A...1G; doi:10.5270/esa-1ug.
Hirose, M., and Osaki, Y. 1990, *Publ. Astron. Soc. Japan*, **42**, 135.
Høg, E., *et al.* 2000, *Astron. Astrophys.*, **355**, L27.
Kato, T. 2015, *Publ. Astron. Soc. Japan*, **67**, 108.
Kennedy, M. R., *et al.* 2016, *Astron. J.*, **152**, 27.
La Dous, C. 1994, *Space Sci. Rev.*, **67**, 1.
Monet, D. G., *et al.* 2003, *Astron. J.*, **125**, 984.
Osaki, Y. 1989, *Publ. Astron. Soc. Japan*, **41**, 1005.
Osaki, Y. 1996, *Publ. Astron. Soc. Pacific*, **108**, 39.
Patterson, J., McGraw, J. T., Coleman, L., and Africano, J. L. 1981, *Astrophys. J.*, **248**, 1067.
Schwarz, G. 1978, *Ann. Statistics*, **6**, 461.
Stellingwerf, R. F. 1978, *Astrophys. J.*, **224**, 953.
Sterken, C. 2005, in *The Light-Time Effect in Astrophysics: Causes and cures of the O–C diagram*, ed. C. Sterken, ASP Conf. Ser. 335, Astronomical Society of the Pacific, San Francisco, 3.
Takeuchi, T. T. 2000, *Astrophys. Space Sci.*, **271**, 213.
Takeuchi, T. T., Yoshikawa, K., and Ishii, T. T. 2000, *Astrophys. J., Suppl. Ser.*, **129**, 1.
Tanabe, K., Akazawa, H., and Fukuda, N. 2018, *Inf. Bur. Var. Stars*, No. 6251, 1.
Vogt, N. 1980, *Astron. Astrophys.*, **88**, 66.

# Appendix A

## A1. Observation log

We show the observation log of our observation of MASTER J004527 in Table A1.

Table A1. Observation log.

| Date (2020) | Start (JST) | End (JST) | Number of Images |
|---|---|---|---|
| 24 Oct | 28:51 | 29:05 | 14 |
| 25 Oct | 19:29 | 29:10 | 468 |
| 26 Oct | 20:01 | 28:43 | 286 |
| 27 Oct | 21:27 | 29:04 | 360 |
| 29 Oct | 19:53 | 28:56 | 342 |
| 30 Oct | 18:59 | 29:18 | 598 |
| 31 Oct | 19:33 | 29:49 | 532 |
| 03 Nov | 18:32 | 20:54 | 142 |
| 04 Nov | 18:07 | 29:21 | 636 |

## A2. AIC and BIC

We first introduce the Akaike information criterion (AIC; Akaike 1974) and the Bayesian information criterion (BIC; Schwarz 1978) formally in the context of maximum likelihood estimation. Let $\ln \mathcal{L}(\theta | \{m_i : i = 1, ...., n\})$ be the log-likelihood where i is the number of photometric observations, $m_i$ is the magnitude observed at time $t_i$, $\{\theta\} = (\theta_1, ...., \theta_k)$ denotes the parameters, and k is the number of parameters. In the classical maximum log-likelihood estimation, we search a set of parameters $\hat{\theta}$ that maximizes $\ln \mathcal{L}(\theta)$ under observed $\{t_i\}$. If we denote the maximum log-likelihood as $\ln \mathcal{L}_{max} \equiv \mathcal{L}(\hat{\theta})$, the AIC is generally defined as

$$\text{AIC} \equiv -2(\ln \mathcal{L}_{max} - k). \quad (A1)$$

Similarly, the BIC is defined as

$$\text{BIC} \equiv -2 \left( \text{Ln } \mathcal{L}_{max} - \frac{k}{2} \ln n \right). \quad (A2)$$

A derivation geared to astronomers is given by Takeuchi (2000).

In the current work, we assumed a polynomial function to describe the shape of humps. Let $t_i$ be the magnitude observed at time $t_i$. To describe the shape of the humps around the peak, we assume a polynomial model as

$$m_i = a_0 + a_1 t_i + a_2 t_i^2 + ... + a_k t_i^k + \epsilon \equiv f(t_i | \{a_k\}) + \epsilon, \quad (A3)$$

where $\epsilon$ is a Gaussian noise with mean 0 and dispersion $\sigma^2$. We consider second and third order (i.e., k = 2 and 3). In order to judge which of the second and third order polynomial models describes the data better with taking into account the penalty of the increase of model parameters, we adopt Akaike's information criterion (AIC; Akaike 1974) and the Bayesian information criterion (BIC; Schwarz 1978). Under the assumption of Equation A3, the AIC and BIC become

Table A2. Estimated AICs and BICs for the polynomial fit.

| Date (JST) | 2nd AIC | BIC | Time of Maxima (JD–2459147) | Error | 3rd AIC | BIC | Time of Maxima (JD–2459147) | Error |
|---|---|---|---|---|---|---|---|---|
| 26 Oct 2020 | –334.6 | –340.7 | 1.9912 | 0.0004 | –344.9 | –339.1 | 1.9926 | 0.0009 |
| 27 Oct 2020 | –318.1 | –314.2 | 3.0393 | 0.0006 | –319.8 | –313.9 | 3.0381 | 0.0026 |
|  | –271.1 | –267.7 | 3.1982 | 0.0004 | –298.1 | –293.1 | 3.1996 | 0.0008 |
|  | –231.3 | –227.8 | 3.2795 | 0.0007 | –230.8 | –225.5 | 3.2797 | 0.0021 |
| 30 Oct 2020 | –370.9 | –367.1 | 6.0110 | 0.0005 | –370.1 | –364.3 | 6.0105 | 0.0039 |
|  | –330.1 | –325.8 | 6.0845 | 0.0011 | –356.0 | –349.5 | 6.0870 | 0.0131 |
|  | –320.5 | –317.0 | 6.1679 | 0.0005 | –318.6 | –313.3 | 6.1679 | 0.0005 |
|  | –297.2 | –293.0 | 6.2524 | 0.0016 | –306.6 | –300.2 | 6.2534 | 0.0101 |

$$\text{AIC}(k) = n \ln \left\{ \frac{\sum_i [m_i - f(t_i | \{\hat{a}_k\})]^2}{n} \right\} + 2(k+1) + n(\ln 2\pi + 1), \quad (A4)$$

$$\text{BIC}(k) = n \ln \left\{ \frac{\sum_i [m_i - f(t_i | \{\hat{a}_k\})]^2}{n} \right\} + (k+1) \ln n + n(\ln 2\pi + 1), \quad (A5)$$

(for a derivation, see, e.g., Takeuchi *et al.* 2000; Banks and Joyner 2017). In practice, the last term $n(\ln 2\pi + 1)$ does not affect the evaluation and we can neglect it. The obtained AIC and BIC are tabulated in Table A2.

**A3. Period estimation by PDM**

As we mentioned in the main text, the PDM is widely used for similar studies. It is well known that some lengths of contiguous data are required, in order to have a secure result by the PDM. However, since we have significant gaps in the observations, clearly seen in Figure 1, the PDM is not an ideal method to have a reliable result. Here, just for a comparison, we applied it to a relatively continuous portion of the current data.

Periodic signals can be approximated more sparsely in the Fourier domain. Therefore, extrapolation techniques that impose sparsity in the Fourier domain can successfully reconstruct missing regions (e.g., Cooray *et al.* 2021a, b) and often perform better than interpolation techniques (e.g., Cooray *et al.* 2020).

The result is summarized in Table A3.

Table A3. Period obtained by PDM.

| Date (JST) | Period (JD–2459147) |
|---|---|
| 26 Oct 2020 | 0.0807272764 |
| 27 Oct 2020 | 0.0803726379 |
| 30 Oct 2020 | 0.0796680278 |